\documentclass[prl,twocolumn,preprintnumbers,amsmath,amssymb,showpacs,superscriptaddress]{revtex4} 
\usepackage{epsfig}
\usepackage{amssymb}
\usepackage{amsmath}
\usepackage{bm}
\usepackage{color}
\usepackage{times}
\usepackage[colorlinks,bookmarks=false,citecolor=blue,linkcolor=red,urlcolor=blue]{hyperref}
\newcommand{\be}{\begin{equation}}

\newcommand{\ee}{\end{equation}}
\newcommand{\bea}{\begin{eqnarray}}
\newcommand{\eea}{\end{eqnarray}}

\begin{document}
\title{Equilibration of a Tonks-Girardeau gas following a trap release 
} 

\author{Mario Collura, Spyros Sotiriadis, and Pasquale Calabrese}
\affiliation{Dipartimento di Fisica dell'Universit\`a di Pisa and INFN, 56127 Pisa, Italy}
\begin{abstract}
We study the non-equilibrium dynamics of a Tonks-Girardeau gas released from a parabolic trap to a circle.
We present the exact analytic solution of the many body dynamics and prove that, for large times and in a properly 
defined thermodynamic limit, the reduced density  matrix of any finite subsystem converges to a generalized Gibbs ensemble.    
The equilibration mechanism is expected to be the same for all one-dimensional systems.
\end{abstract}
\maketitle

The non-equilibrium dynamics of isolated many body quantum systems is currently in a 
golden age mainly due to the experiments on trapped ultra-cold atomic gases
\cite{uc,kww-06,tc-07,tetal-11,cetal-12,getal-11,shr-12,rsb-13} in which 
it is possible to measure the unitary non-equilibrium evolution without any significant coupling to the environment. 
A key question is whether the system relaxes to a stationary state, and if it does, 
how to characterize from first principles its physical properties at late times. 
It is commonly believed that, depending on the integrability of the Hamiltonian governing the time evolution, 
the behavior of local observables either can be described by an effective thermal 
distribution or by a generalized Gibbs ensemble (GGE), for non-integrable and integrable systems 
respectively (see e.g. \cite{revq} for a review). 
While this scenario is corroborated by many 
investigations \cite{cc-06,gg,c-06,cdeo-08,bs-08,r-09,CEF,CEFII,f-13,eef-12,se-12,ccss-11,rs-12,bdkm-11,fm-10,sfm-12,mc-12,ce-13,fe-13,gge-new}, 
a few studies \cite{kla-07,bch-11,rf-11,gme-11,gm-11,gp-08} suggest that the 
behavior could be more complicated and in particular can depend on the initial state.

\begin{figure}[t]
\includegraphics[width=0.5\textwidth]{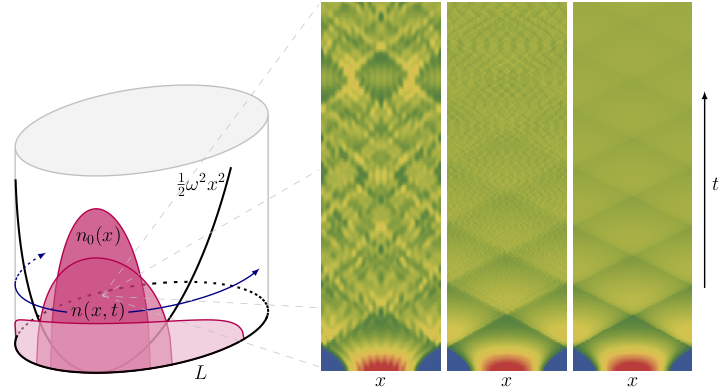}
\caption{Left: sketch of the trap release dynamic in a circle.
Right: Color plot of the numerical calculated density evolution for $N=10,100,\infty$ (from left to right) 
at $N/L=1/2$ and $\omega N=5$.}
\label{sketch}
\end{figure}

In a global quantum quench, the initial condition is the ground state of a translationally invariant Hamiltonian
which differs from the one governing the evolution by an experimentally tunable parameter such as a magnetic field.
A different initial condition can be experimentally achieved \cite{shr-12,rsb-13} by 
considering the non-equilibrium dynamics of a gas released from a parabolic trapping potential. 
It has been shown experimentally that  the spreading of correlations is 
ballistic for an integrable system and diffusive for a non-integrable one \cite{rsb-13}.
However, when the gas expands in  full space, for infinite time the gas clearly reaches zero density 
(see e.g. \cite{mg-05,gp-08,v-12,a-12,cro,hm-v} for a theoretical analysis) and it is rather confusing 
to distinguish thermal and GGE states. 
To circumvent this,  Caux and Konik \cite{ck-12}  have recently 
 developed a new approach based on integrability to 
study the release of the Lieb-Liniger Bose gas \cite{LiebPR130} from a parabolic trap not in  free space but 
on a closed circle (as sketched in Fig. \ref{sketch}),
so that the gas has finite density. 
It has been numerically shown that the time averaged 
correlation functions are described by a GGE, apart from finite size effects \cite{ck-12}.
A preliminary analysis for non-integrable models has also been presented \cite{bck-13}.
However, while this approach permits effectively to calculate time-averaged quantities for relatively large  
systems (the maximum number of particles is $N=56$  \cite{ck-12}), 
the study of the time evolution is possible but much harder and 
it is difficult to establish whether (and in which sense) an infinite time limit exists.

In order to overcome these limitations, we present here a full analytic solution of this non-equilibrium 
dynamics in the limit of strong coupling, i.e. in the celebrated Tonks-Girardeau regime \cite{TG}. 
We will show that, in a properly defined thermodynamic (TD) limit, the reduced density matrix of any 
{\it finite} subsystem converges for long times to the GGE one. 
This implies that any measurable {\it local} observable will converge to the GGE predictions.

{\it  The model and quench protocol}.
We consider a one-dimensional Bose gas with delta pairwise interaction and in an external parabolic potential
with Hamiltonian 
\begin{equation}
H = -\frac12\sum_{j=1}^N \frac{\partial^2}{\partial{x_j^2}}  +\sum_{j=1}^N \frac12 \omega^2  x_j^2+ c \sum_{ i\neq j } \delta(x_i - x_j),
\label{LL}
\end{equation}
where $c > 0$ is the coupling constant (we set $\hbar  = m=1$).  
The translationally invariant Lieb-Liniger model is obtained for $\omega=0$ and
on a circle of length $L$ with periodic boundary conditions (PBC).
While Ref. \cite{ck-12} covers numerically arbitrary $c$, to make an analytic progress we consider  
the strong-coupling limit of impenetrable bosons $c\to\infty$,
corresponding also to the low density $n\equiv N/L\ll1$ regime 
for any $c$ \cite{LiebPR130}.

For a trap release, the initial state is  a Tonks Girardeau gas confined by 
a parabolic potential, i.e. the ground state of Eq. (\ref{LL}) for a fixed $\omega$.
Following \cite{TG}, the many body wave function for the ground state of $N$ impenetrable bosons is
 \begin{equation}
 \Psi_B(x_1,\cdots,x_N)=\prod_{ i < j} {\rm sgn}(x_j-x_i) \Psi_F(x_1,\cdots,x_N),
  \label{AF}
\end{equation}
where $\Psi_F(x_1,\cdots,x_N)$ is the ground-state function of 
$N$ free fermions in the parabolic potential, i.e. the Slater determinant 
$\det_{i,j} \chi_j(x_i)$ with the eigenstates of the harmonic oscillator 
\be
 \chi_j(x) = \frac{1}{\sqrt{2^j j!}} \left(\frac{\omega}{\pi}\right)^{1/4} e^{- {\omega x^2}/2} H_j\left(\sqrt\omega x \right), 
 \ee
and $H_j(z)$ the Hermite polynomials.
In this fermionic language, the time evolution governed by the Hamitonian (\ref{LL}) with $\omega=0$ 
is obtained by expanding the one-particle states in the 
free-wave basis, i.e. ($k=2\pi m/L$)
\be
\chi_j(x)= \sum_{k} A_{k,j}\frac{e^{-i k x}}{\sqrt{L}},\quad A_{k,j}= 
\int_{-\frac{L}2}^{\frac{L}{2}} \hspace{-1mm} dx  \chi_j(x) \frac{e^{i k x}}{\sqrt{L}}. 
\label{overlap}
\ee
We now make the only crucial {\it physical} assumption:
we impose that the space initially occupied by the trapped gas as a whole is  within the external box of length $L$, i.e. 
before the quench the PBC are irrelevant for the gas which only ``sees'' the parabolic trap. 
This condition is what allows us to talk about {\it release} of the gas
and requires the number of particles $N$ to be smaller than the first level  
of the parabolic potential that is affected by PBC. 
In the TD limit, for large quantum numbers, $|\chi_N(x)|^2$ is the semiclassical probability density at the 
corresponding energy that tends to zero for $|x|>\ell/2$
with $\ell$ the classical cloud dimension
$\ell = 2\sqrt{2N/\omega}$.  
In simpler words, this means that the classical extension of the gas in the trap $\ell$ must be smaller 
than the box size $L$.

To have a well-defined TD limit, we should consider $N,L\to\infty$
at fixed density $n\equiv N/L$ and, at the same time, 
$\omega\to0$
with $\omega N$ constant (fixed initial density), as in \cite{ck-12}.
In terms of these quantities the gas release condition $\ell<L$ reads 
$\sqrt{N\omega} > 2\sqrt2 n$ and
the coefficients $A_{k,j}$ can be calculated extending the integration in Eq. (\ref{overlap}) to $\pm\infty$,
obtaining 
\be
 A_{k,j} = i^j \sqrt{\frac{2\pi}{\omega L}} \chi_j(k/\omega). 
 \label{Akj}
\ee
Also the infinite time limit should be handled with care. Indeed, in  this quench, 
a stationary behavior is possible because of the interference of  the particles going around the circle $L$
many times (see Fig. \ref{sketch}), i.e. to observe a stationary value we must require $vt\gg L$ (the speed of sound is  
$v=\sqrt{2\omega N}$ in our normalization). 
This is very different from equilibration in standard global
quenches where the time should be such that the boundaries are not reached (see e.g. \cite{CEFII}) in order to 
avoid revival effects. In this problem the revival scale is $t_{\rm r}\propto L^2$ and so 
the infinite time limit in which a stationary behavior can be achieved  is 
$t/L\to \infty$ provided  $t/L^2\to 0$. 
The importance of the TD and long time limits to get a stationary behavior is already evident from 
the time evolution of the density profile in Fig. \ref{sketch}.

\begin{figure}[t]
\includegraphics[width=0.25\textwidth]{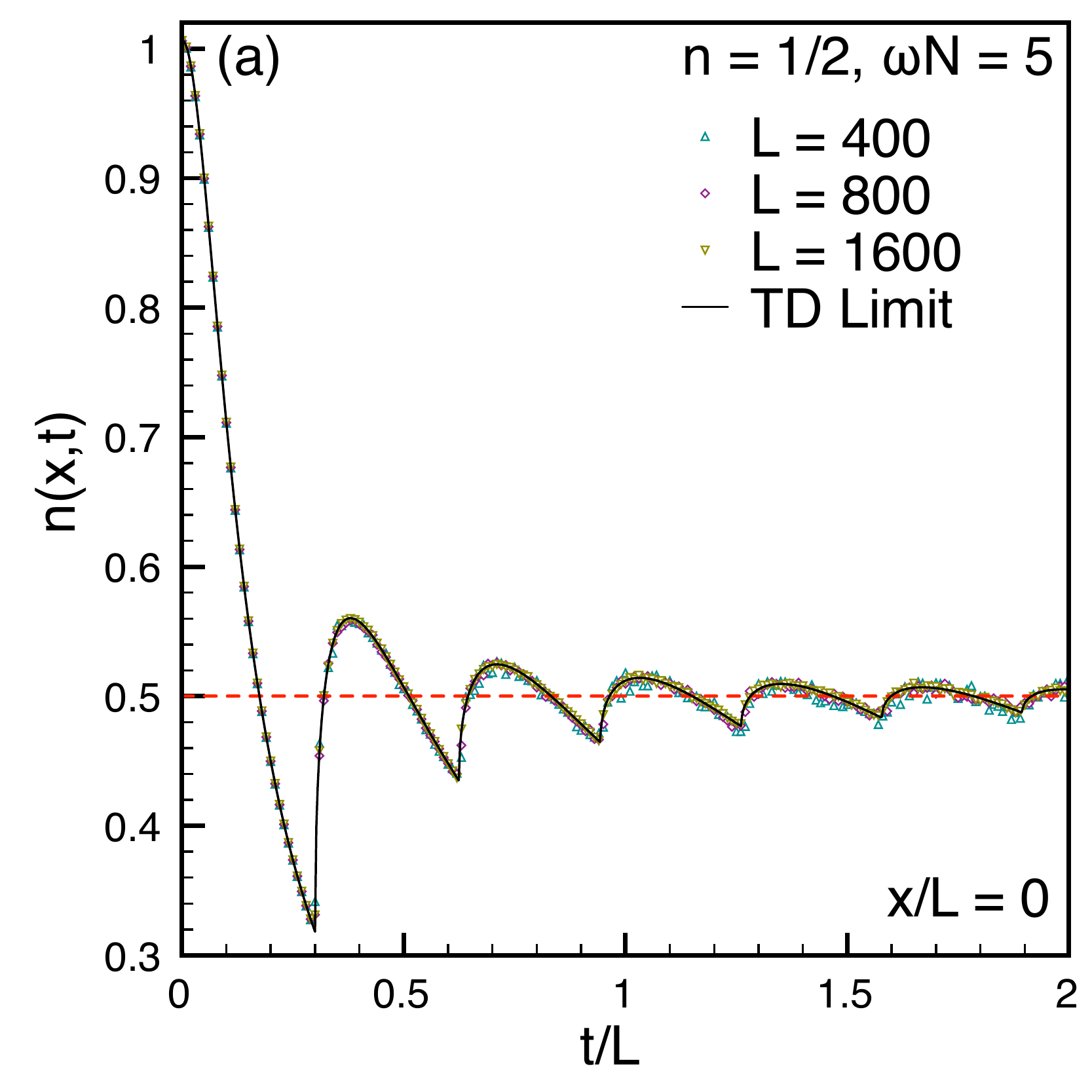}\includegraphics[width=0.25\textwidth]{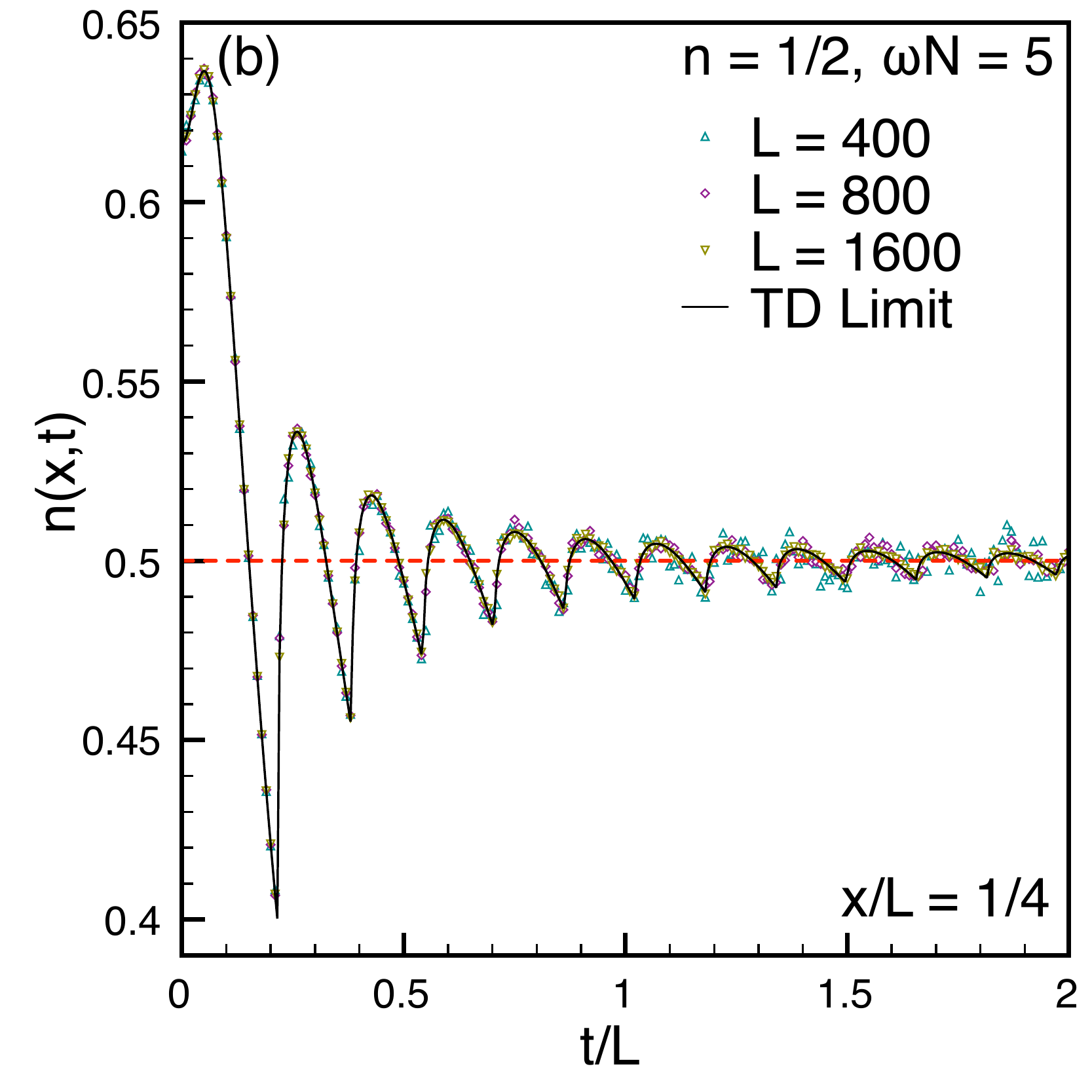}\\\includegraphics[width=0.25\textwidth]{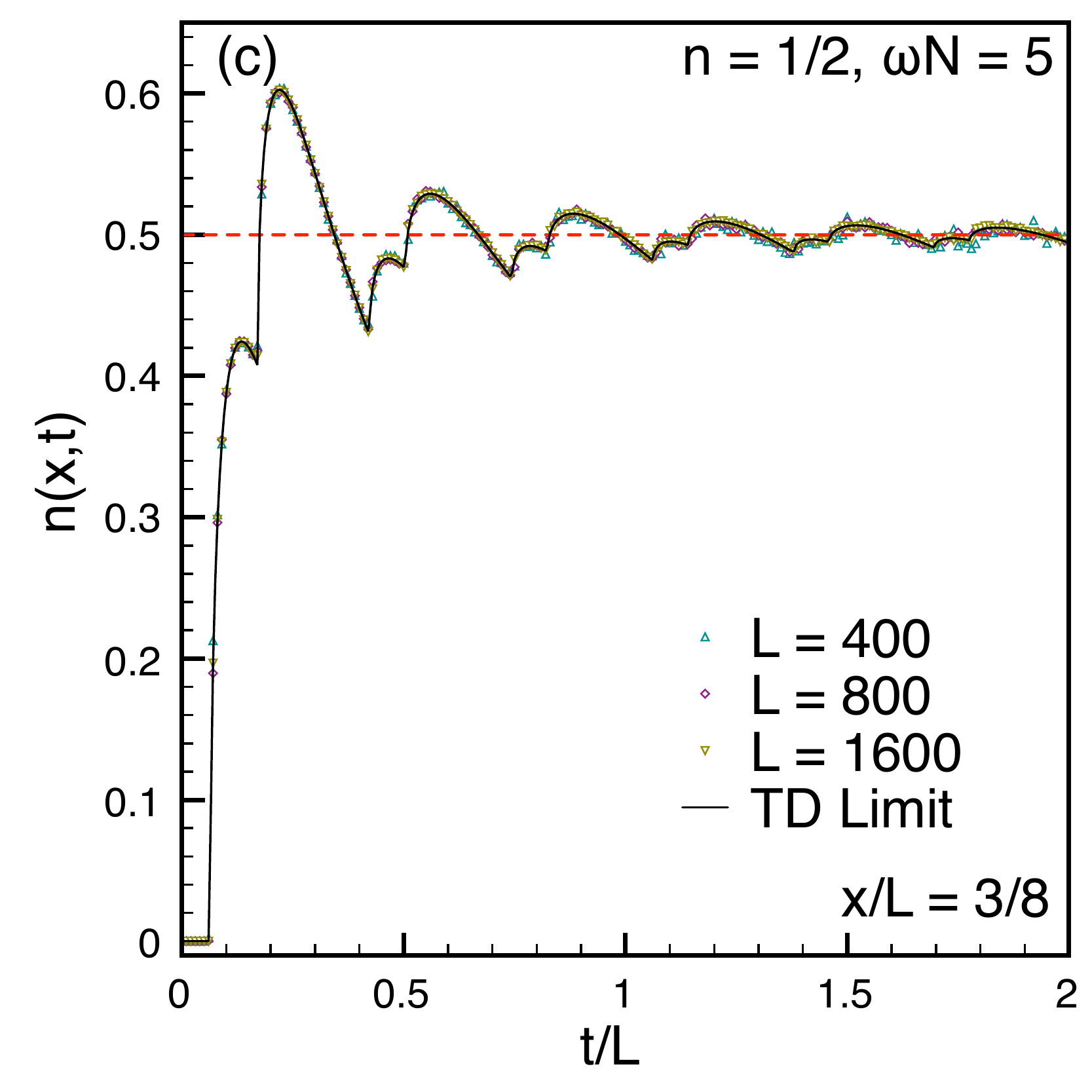}\includegraphics[width=0.25\textwidth]{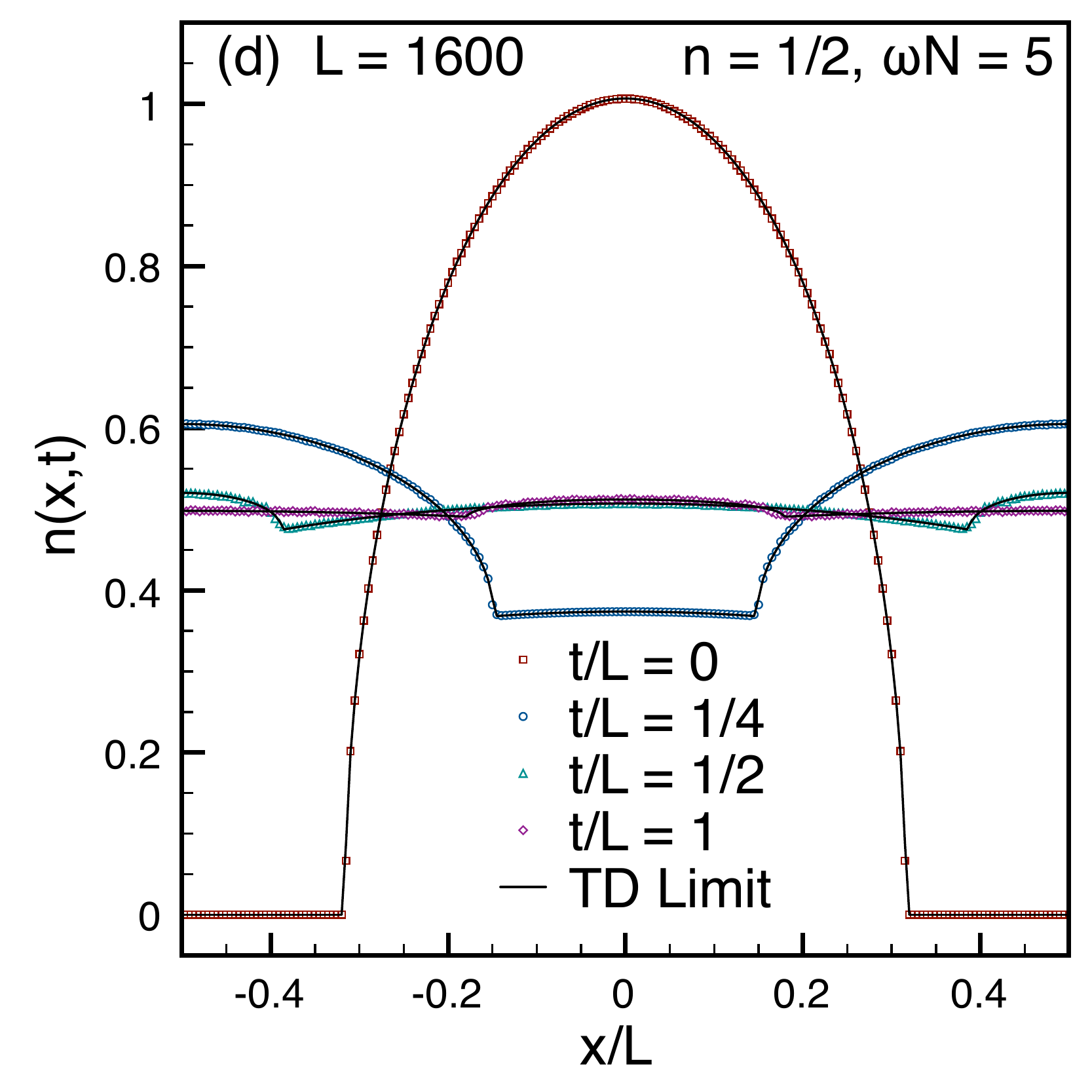}

\caption{(a,b,c) Time evolution of the density $n(x,t)$ for different $x/L$ and sizes. 
Dashed red lines indicate the equilibration value $N/L$ at infinite time. 
(d)  Density profile for $L=1600$ at different rescaled times $t/L$. 
Symbols are the exact dynamics for finite $N$, while full black lines are the TD limit.} 
\label{figden}
\end{figure}

{\it The one-particle problem}. In fermion language, the time dependent many body  state
is the Slater determinant of the time evolved one-particle initial eigenfunctions (i.e. the solution of the 
Schrodinger equation $i\partial_t \Phi_j(x,t)= H_F \Phi_j(x,t)$ with $\Phi_j(x,0)=\chi_j(x)$ and $H_F$ the
single particle free Hamiltonian with PBC).
These time evolved wave functions can be calculated from Eqs. (\ref{overlap}) and (\ref{Akj}) obtaining 
\begin{equation}
\Phi_j(x,t) = \sum_{p=-\infty}^{+\infty} {\Phi^\infty_j}(x+pL,t),
\label{Phisum}
\end{equation}
where 
\begin{multline}
\Phi_j^\infty(x,t) = 
 \frac{(1-i\omega t)^{\frac{j}2}e^{-i \frac{t\omega^2  x^2}{2(1+\omega^2 t^2)}}}{(1+i\omega t)^{\frac{j+1}2}}  
\chi_{j}\Big(\frac{x}{\sqrt{1+\omega^2 t^2}}\Big),
\label{Phixt}
\end{multline}
is the time evolved eigenfunction in infinite space which agrees with the result in \cite{mg-05}.
The boson-fermion mapping remains valid for the time-dependent problem \cite{gw-00}.

{\it Time evolution of the density profile.} 
We start the analysis of the many body problem from the density profile $n(x,t)= \sum_j |\Phi_j(x,t)|^2$ 
which shows clearly how a non-zero stationary value can be achieved in a trap release experiment. 
From Eq. (\ref{Phixt}) we have for arbitrary time, $N,L,\omega$
\begin{multline}
n(x,t) = 
\frac{1}{\sqrt{1+\omega^2 t^2}}  
\sum_{p,q=-\infty}^{\infty} \hspace{-1mm} e^{ i\frac{\omega^2 t}{2(1+\omega^2 t^2)}[(x+pL)^2 - (x+qL)^2 ]} \\ \times
\sum_{j=0}^{N-1} \chi_{j}\left( \frac{x+pL}{\sqrt{1+\omega^2 t^2}}\right) \chi_{j}\left( \frac{x+qL}{\sqrt{1+\omega^2 t^2}}\right),
\end{multline}
which, in the TD limit, because of the strongly oscillating phase factor, reduces to the diagonal part $p=q$: 
\begin{equation}
n(x,t) = \frac{1}{\sqrt{1+\omega^2 t^2}}  \sum_{p=-\infty}^{\infty} \sum_{j=0}^{N-1} 
\left|\chi_{j}\Big( \frac{x+pL}{\sqrt{1+\omega^2 t^2}}\Big)\right|^2, 
\end{equation}
and can be rewritten in terms of the TD limit of the particle density at initial time $n_0(x)=(\sqrt{2N \omega-\omega^2x^2})/{\pi}$
 as
\begin{equation}
n(x,t) = \frac{1}{\sqrt{1+\omega^2 t^2}} \sum_{p=-\infty}^{\infty} n_{0}\left(\frac{x+pL}{\sqrt{1+\omega^2 t^2}}\right),
\end{equation}
In Figs. \ref{sketch} and \ref{figden} we show the numerically calculated time dependent density for finite but large $N$
which perfectly agrees with the above TD prediction for any time.

\begin{figure}[t]
\includegraphics[width=0.25\textwidth]{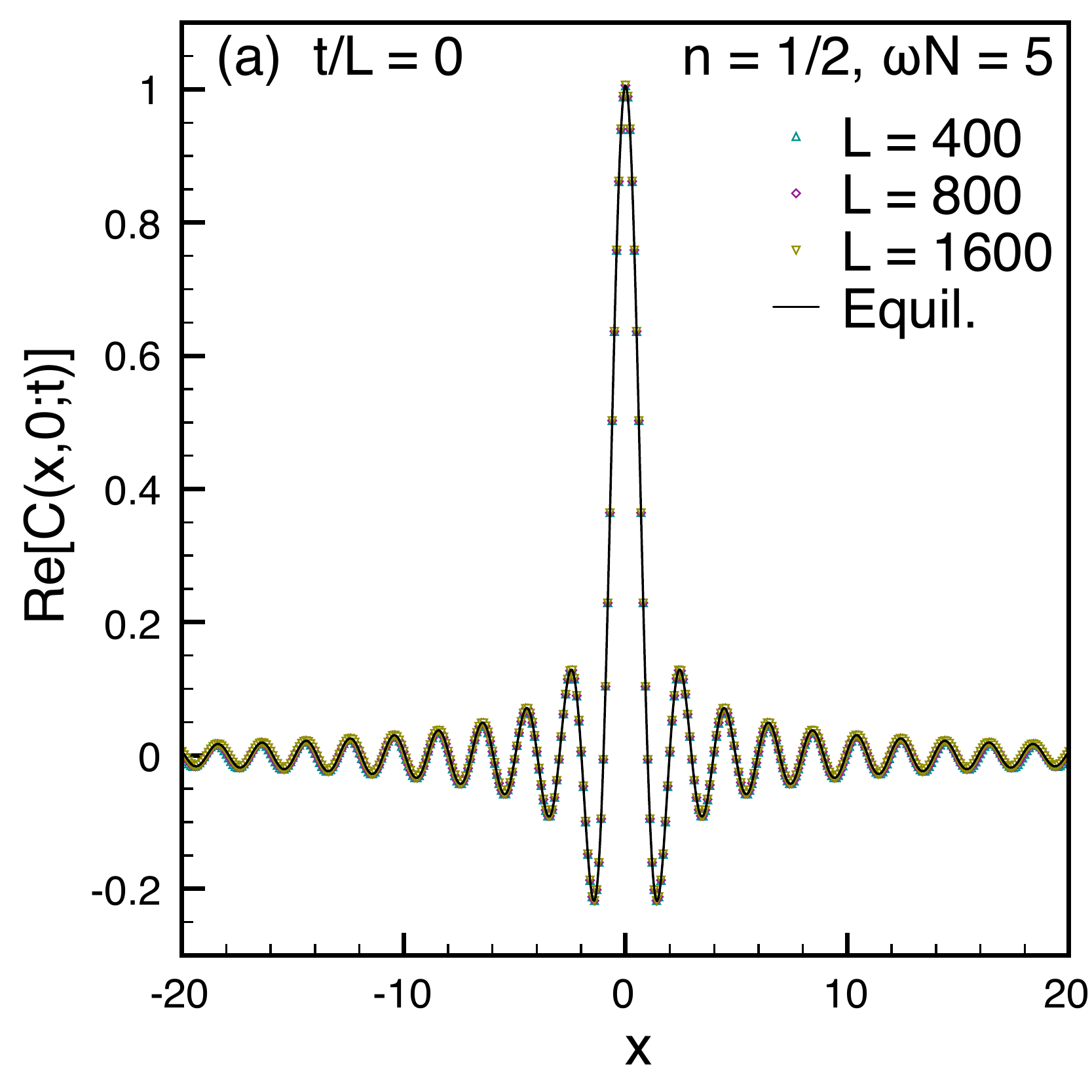}\includegraphics[width=0.25\textwidth]{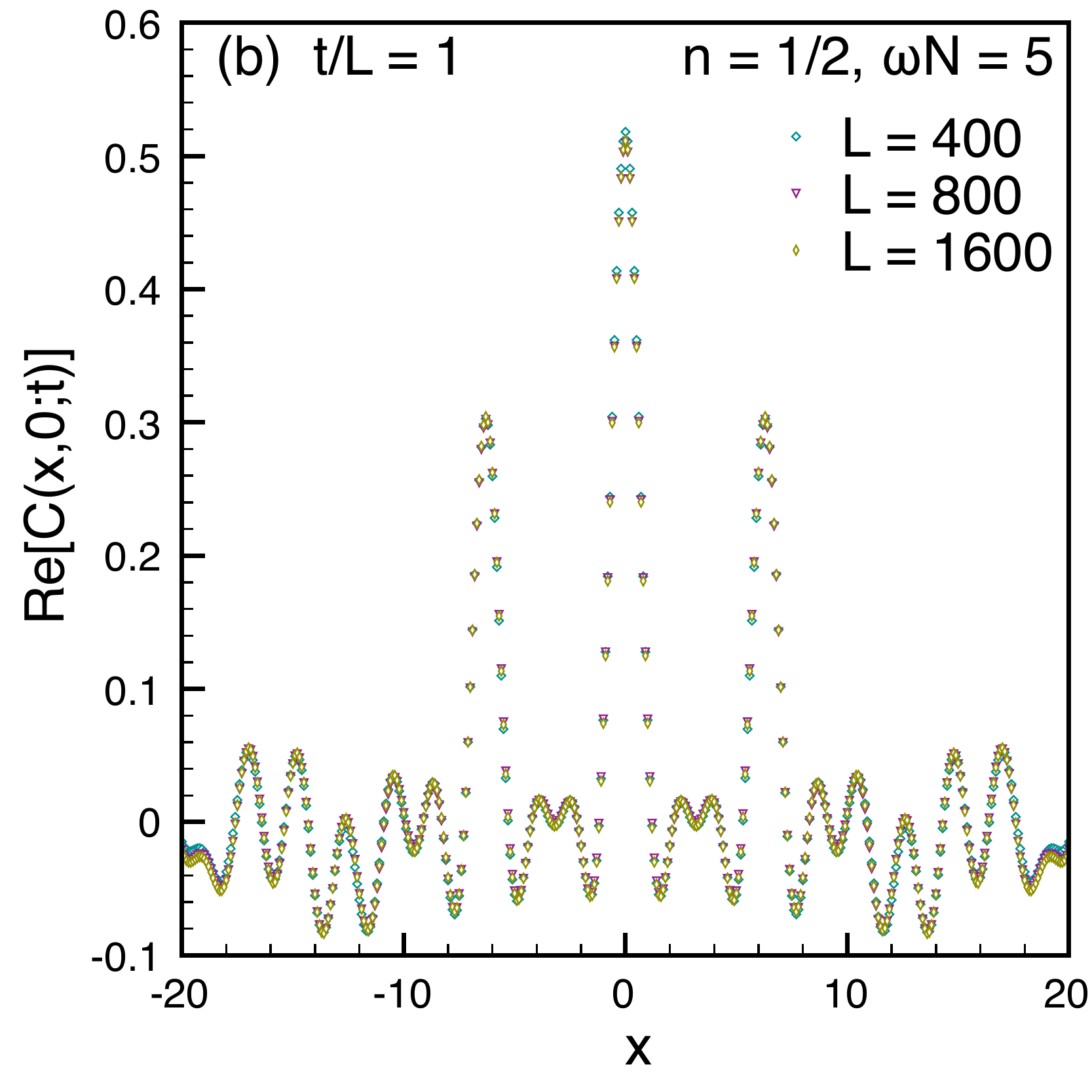}\\
\includegraphics[width=0.25\textwidth]{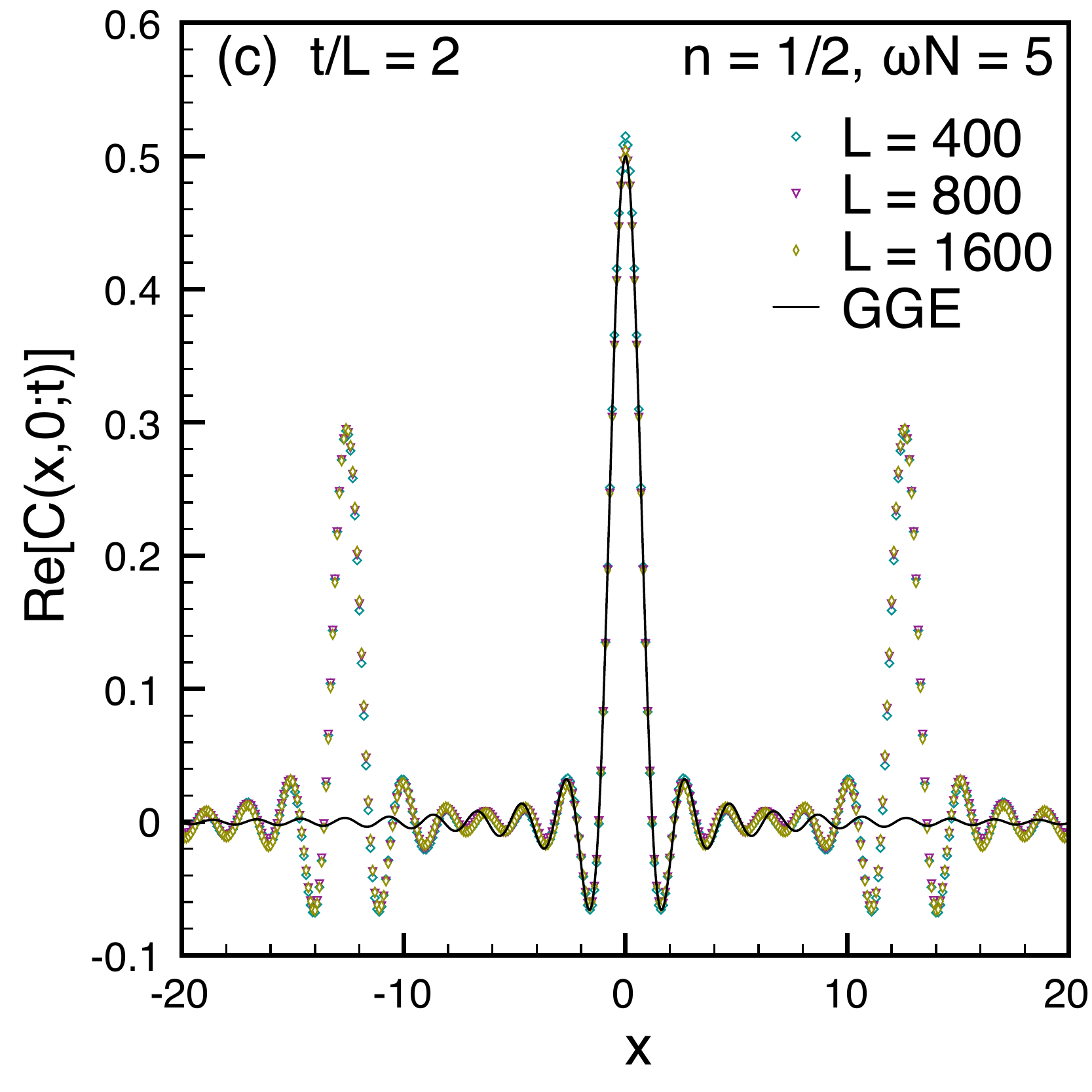}\includegraphics[width=0.25\textwidth]{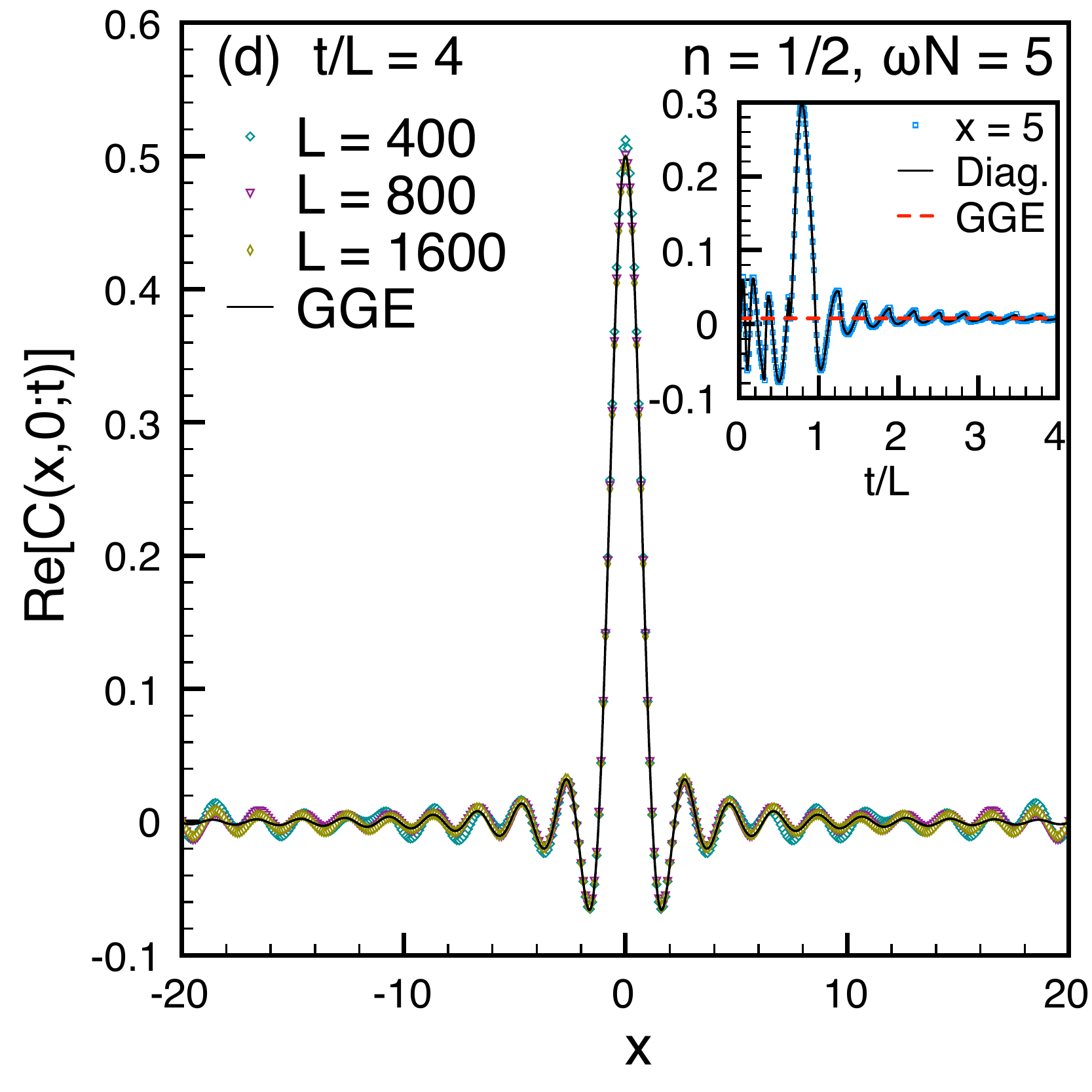}

\caption{Snapshots of the correlation ${\rm Re}[C(x,0;t)]$ at different rescaled times $t/L$ and sizes. 
For $t/L=0$ the full line is the initial correlation in the TD limit, i.e. $C(x,0)=\sin[\sqrt{2\omega N}x]/\pi x$ valid for 
$x\ll L$. 
The full line for $t/L=2,4$ is the stationary value in Eq. (\ref{Cxyinfty}). 
As time increases, two symmetric peaks are {\it expelled} from the central region. 
The inset in (d) shows the evolution for fixed $x=5$ and $L=1600$: 
after  the moving peak has been expelled, 
the correlation is damped in time and converges to the GGE. 
} 
\label{figCxy}
\end{figure}

{\it The two-point fermionic correlator} 
$C(x,y;t)\equiv\langle \Psi^{\dag}(x,t) \Psi(y,t) \rangle$ is given by
\begin{equation}
C(x,y;t)= \sum_{j=0}^{N-1} \Phi^*_{j}(x,t)\Phi_{j}(y,t).
\end{equation}
The numerical determination of this correlation function for finite $N$ is reported in Fig. \ref{figCxy} showing the approach to the  
infinite time limit \cite{SM}
\begin{equation}
C(x,y; t\to\infty) = 2 n \frac{J_1[\sqrt{2\omega N} (x-y)]}{\sqrt{2\omega N}(x-y)},
\label{Cxyinfty}
\end{equation}
with $J_1(z)$ the Bessel function. 

{\it The reduced density matrix and the GGE.}
For a closed system evolving under Hamiltonian dynamics, the existence of a stationary state 
may seem paradoxical  because the whole  system is always in a pure 
state and cannot be described by a mixed state at infinite time.
This `paradox' is solved in the reduced density matrix formalism:
given a space interval $A$, the  reduced density matrix is $\rho_A(t)={\rm Tr}_B \rho(t)$ where
$B$ is the complement of $A$ and $\rho(t)=|\Psi(t)\rangle\langle \Psi(t)|$ is the density matrix of the 
whole system. 
With some abuse of language, we say that a system becomes stationary if, after the TD limit is taken for the whole system, 
the limit $\displaystyle\rho_{A,\infty}=\lim_{t\to\infty} \rho_A(t)$ exists for any finite $A$  \cite{CEFII}.
Furthermore we say that a system is described by a statistical ensemble with density matrix $\rho_E$
if the reduced density matrix $\rho_{A,E}\equiv {\rm Tr}_B \rho_E$ equals $\rho_{A,\infty}$.

For a gas of free fermions, by means of Wick theorem, any observable can be obtained from the two-point correlator. 
The construction of $\rho_A$ in terms of $C(x,y)$ in continuous space has been detailed in \cite{cmv-11} 
(generalizing the lattice approach \cite{pes}).
As a crucial point, the non-local transformation mapping the Tonks-Girardeau gas to 
free fermions is local within any given {\it compact} subspace, 
i.e. the bosonic degrees of freedom within $A$ can be written only in terms of fermions in $A$.
This is analogous to lattice models such as the Ising chain \cite{CEF,CEFII,f-13}. 
Thus, if  for finite $x,y$, $C(x,y,t\to\infty)$ is described by a statistical ensemble, also $\rho_A$ will be
and consequently  the expectation value of any  observable local within $A$.

Because of integrability, it is natural to expect that Eq. (\ref{Cxyinfty}) should be 
described by a GGE
\be
\rho_{GGE}= Z^{-1} e^{-\sum \lambda_i I_i}, 
\ee
with $\{I_i\}$ a complete set of {\it local} integrals of motion and $\lambda_i$ 
Lagrange multipliers fixed by the conditions $\langle \Psi_0| I_i |\Psi_0\rangle={\rm Tr} [\rho_{GGE}I_i]$, with $|\Psi_0\rangle$
the many body initial state.
However, for free fermions, 
one can work with the momentum occupation modes $\hat n_k=c^\dagger_k c_k$ 
which are non-local integrals of motion, but can be written as linear combinations of local integrals of motion \cite{fe-13}. 
In the TD limit, the initial values of $\hat n_k$ are 
\be
\langle \Psi_0| \hat n_k |\Psi_0\rangle =\sum_{j=0}^{N-1} |A_{k,j}|^2\simeq 
\frac{2}{L} \sqrt{\frac{2N}{\omega}} \sqrt{1-\frac{k^2}{2\omega N}},
\label{nkGGE}
\ee
and zero if the argument of the square root is negative. 
In the GGE we have ${\rm Tr} [\rho_{GGE} \hat n_k]= (e^{\lambda_k}+1)^{-1}$ 
and equating the two,  the $\lambda_k$ are derived.
It is now straightforward to show that $C(x,y)$ in the GGE equals the infinite time limit 
of trap release in Eq. (\ref{Cxyinfty}) \cite{SM}.
This shows that all stationary quantities of the released gas are described by a GGE.
Furthermore, in Ref. \cite{eef-12} it has been shown that all 
non-equal time stationary properties are always determined by the same ensemble describing the static quantities,
and so, even in our case, they are encoded solely in the GGE.

\begin{figure}[t]
\includegraphics[width=0.4\textwidth]{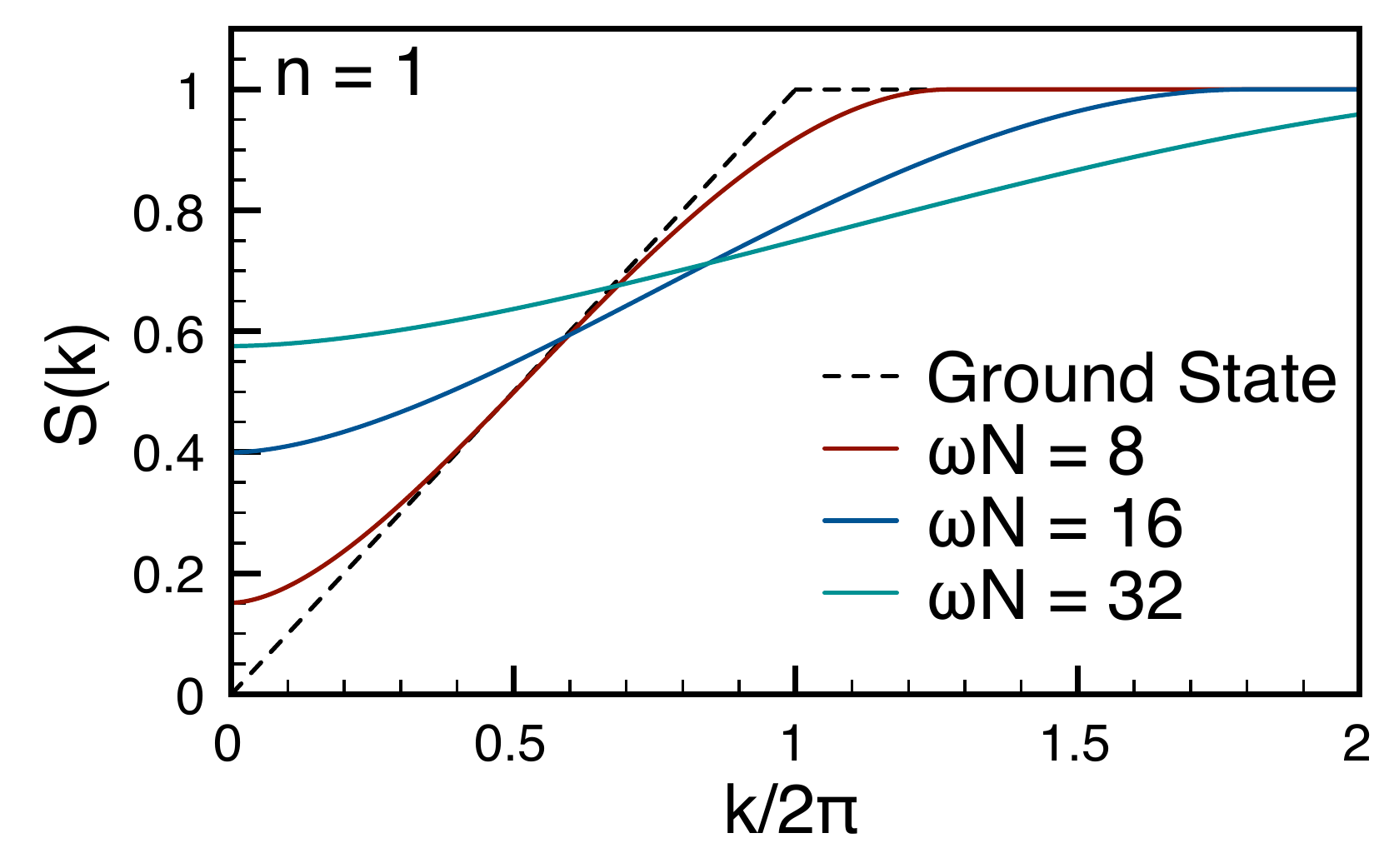}
\caption{The GGE structure factor $S(k)$ as function of $k/2k_F$ ($k_F=\pi n$) for different initial
trap potentials $\omega N$  compared with the ground-state one (dashed line).
} 
\label{figSk}
\end{figure}

{\it The structure factor in the GGE.}
The structure factor $S(k)$ is the Fourier transform of the density-density correlation
$\langle \hat n(x,t) \hat n(0,t)\rangle$.  
In any ensemble which is diagonal in the Fourier modes, in the TD limit the structure 
factor can be written in terms of occupation modes $n_k$ as
\be
1-S(k)=  \frac L{N} {\int} \frac{dq}{2\pi} \; n_q n_{k-q} 
= \frac{4 \sqrt2 n}{\pi \sqrt{\omega N}} f\Big(\frac{k}{\sqrt{2\omega N}}\Big),
\ee
where the rhs is obtained using the GGE $n_k$ given in Eq. (\ref{nkGGE}). 
Here  $f(x)= [(4 + x^2) E(1 - {4}/{x^2}) - 8 K(1 - {4}/{x^2})] {|x|}/6$ if $|x|<2$
and zero otherwise where
 $E(z)$ and $K(z)$ are standard elliptic functions and $f(0)=4/3$. 
$S(k)$ turns out to be an even function of $k$ and monotonic for $k>0$. 
The plot of $S(k)$ for different  initial trapping potentials is reported in Fig. \ref{figSk}. 
$S(k)$ resembles the one found numerically in \cite{ck-12} for the Lieb-Liniger gas. 
Because of the trap release constraint $\sqrt{N\omega} > 2\sqrt2 n$, we have
$S(k)>S(0)\geq 1-8/3\pi=0.151174\dots$.
This calculation shows how easy it is to obtain GGE predictions without solving the full 
non-equilibrium dynamics.

{\it The bosonic two-point function} or one-body density matrix 
$C_{B}(x,y;t)\equiv \langle\hat{\Phi}^\dagger(x,t)\hat{\Phi}(y,t) \rangle$ (with $\hat{\Phi}(y,t)$ bosonic annihilation operator)
is a non-trivial quantity whose calculation presents difficulties also in thermal equilibrium \cite{vm-13}. 
However, using the approach in \cite{pezer}, the computation is easy 
for large time and in the TD limit obtaining \cite{SM} 
\be
C_{B}(x,y;t\rightarrow\infty)  =  C(x,y;t\rightarrow\infty) e^{-2n|x-y|}, 
\ee
with $C(x,y;t\to\infty)$ the fermion correlator in Eq. (\ref{Cxyinfty}).
For small distances, $C_{B}(x,y;t\rightarrow\infty)$  shows a singular behavior of the form $|x-y|$
which is different from its thermal counterpart $|x-y|^3$ \cite{vm-13}.
This behavior is strictly valid only in the TD limit because for any finite $N$, 
at very small distances $C_{B}(x,y;t\rightarrow\infty)$ crosses over to $|x-y|^3$
as expected from general arguments \cite{vm-13}. 
This finite $N$ crossover is numerically demonstrated in \cite{SM}.
Consequently, the momentum distribution function has a large momentum tail of the form $k^{-2}$
which crosses over to the standard $k^{-4}$ for even larger $k$. 
This large-momentum crossover should be a measurable signature of the GGE.

{\it Trap to trap release}.
The case of  a Tonks-Girardeau gas released not in a periodic system but in a larger harmonic trap 
has been solved by Minguzzi and Gangardt \cite{mg-05} who
showed that the system oscillates forever without relaxation. 
However, even in this case, it is simple to see that the time averaged two-point correlations 
(and hence by Wick theorem any observable) are still described by a GGE.

{\it Conclusions}. 
In this letter we solved analytically the non equilibrium  dynamics of a Tonks Girardeau gas
following a trap release to a periodic geometry as in Fig. \ref{sketch}. 
We prove that for long time and in the TD limit,  any finite subsystem becomes stationary and its behavior 
is described by a GGE. This provides the first analytic proof of a GGE for an inhomogeneous initial state. 
We stress that the mechanism responsible for the equilibration is very different from the one in a global quantum 
quench since in the trap release it is due to the interference of the particles going around the circle many times.
This equilibration mechanism is expected to be the same for any one dimensional gas released into a circle. 

Apart from the per se experimental interest \cite{shr-12,rsb-13}, these results represent a first step 
towards a complete analytical understanding of the famous quantum Newton cradle \cite{kww-06} 
at least in the Tonks Girardeau limit. 

{\it Acknowledgments}. 
We are grateful to F. Essler, M. Kormos, and E. Vicari  for helpful discussions.  
All authors  acknowledge the ERC  for financial  support under  Starting Grant 279391 EDEQS.

\newpage

\clearpage 
\setcounter{equation}{0}%
\renewcommand{\theequation}{S\arabic{equation}}
\setcounter{page}{1}

\onecolumngrid

\section{Supplementary Material}

\subsection{The two-point fermionic correlation function}
Plugging the one-particle time-evolved wavefunctions [Eqs. (6) and (7) in the main text] in the definition of the 
fermionic correlator $C(x,y;t) = \sum \Phi^{*}_{j}(x,t) \Phi_{j}(y,t)$, we have 
\bea\label{Cxyt_1}
C(x,y;t) &=& \frac{1}{\sqrt{1+\omega^{2}t^2}}\sum_{p,q=-\infty}^{\infty}\exp\left\{i \frac{\omega^2 t}{2(1+\omega^2 t^2)} [(x+pL)^2-(y+qL)^2] \right\}\nonumber\\
&\times& \sum_{j=0}^{N-1}\chi_{j}\left(\frac{x+pL}{\sqrt{1+\omega^{2}t^2}}\right)\chi_{j}\left(\frac{y+qL}{\sqrt{1+\omega^{2}t^2}}\right).
\eea
Due to the oscillating phase factor, in the TD limit, the leading behavior of Eq. (\ref{Cxyt_1}) is given only by the 
diagonal part  $p=q$, so that 
\be\label{Cxyt_2}
C(x,y;t) \simeq \frac{\mathrm{e}^{i \frac{\omega^2 t(x^2 - y^2)}{2(1+\omega^2 t^2)}}}{\sqrt{1+\omega^{2}t^2}}\sum_{p=-\infty}^{\infty}\exp\left\{i \frac{\omega^2 t(x-y)pL}{1+\omega^2 t^2} \right\}
\sum_{j=0}^{N-1}\chi_{j}\left(\frac{x+pL}{\sqrt{1+\omega^{2}t^2}}\right)\chi_{j}\left(\frac{y+pL}{\sqrt{1+\omega^{2}t^2}}\right),
\ee
that in the regime $t\gg \omega^{-1}$ gives
\bea\label{Cxyt_larget}
C(x,y;t) &\simeq& \frac{1}{\omega t}\sum_{p=-\infty}^{\infty}\mathrm{e}^{i (x-y)pL/t}
\sum_{j=0}^{N-1}\chi_{j}\left(\frac{x+pL}{\omega t}\right)\chi_{j}\left(\frac{y+pL}{\omega t}\right)\\
&\label{Cxyt_larget2} = &
\frac{1}{\sqrt{\omega} t}\sum_{p=-\infty}^{\infty}\mathrm{e}^{i (x-y)pL/t}
\sum_{j=0}^{N-1}\tilde{\chi}_{j}\left(\frac{x+pL}{\sqrt{\omega} t}\right)\tilde{\chi}_{j}\left(\frac{y+pL}{\sqrt{\omega} t}\right).
\eea
where we defined $\tilde{\chi}_{j}(x)$ as the eigenfunctions of 
a harmonic oscillator with $\omega=1$ (i.e. $\tilde{\chi}_{j}(x)={\chi}_{j}(x)|_{\omega=1}$).

In the large time limit $t/L\to \infty$ (but with $t/L^{2}\rightarrow 0$), the sum over $p$ can be replaced by an integral
\be
 \sum_{p=-\infty}^{\infty} \gamma F( \gamma p )\longrightarrow \int_{-\infty}^\infty dz F(z),
\ee
where $\gamma=L/(\sqrt{\omega}t)\to 0$, allowing to rewrite $C(x,y;t)$ as
\bea\label{Cxyt_larget3}
C(x,y;t\rightarrow\infty)& = & \frac{1}{L}\int_{-\infty}^{\infty} dz\,\mathrm{e}^{i\sqrt{\omega} (x-y)z}
\sum_{j=0}^{N-1}\tilde{\chi}_{j}\left(\gamma\frac{ x}{L}+z\right)\tilde{\chi}_{j}\left(\gamma\frac{ y}{L}+z\right) \\
& = &  \frac{1}{L}\int_{-\infty}^{\infty} dz\,\mathrm{e}^{i\sqrt{\omega} (x-y)z}
\sum_{j=0}^{N-1} \left | \tilde{\chi}_{j}(z)\right|^{2},
\eea
where in the second line we used once again $\gamma\to0$.
Now using
\be
\lim_{N\to\infty} \left(\sum_{j=0}^{N-1} \left | \tilde{\chi}_{j}(x)\right|^{2}\right)= \frac{\sqrt{2N-x^2}}\pi\equiv \tilde{n}_{0}(x) ,
\ee
we have in the TD limit
\be\label{Cxyt_larget_leading}
C(x,y;t\to\infty)  =  
\frac{1}{L}\int_{-\infty}^{\infty}dk \mathrm{e}^{i\sqrt{\omega}\,(x-y)k} \tilde{n}_{0}(k) =
\int_{-\infty}^{\infty}\frac{dk}{2\pi} \mathrm{e}^{i(x-y)k} \langle \Psi_0| \hat n_k|\Psi_0\rangle=
  2n\frac{J_{1}\left[ \sqrt{2\omega N}(x-y) \right]}{\sqrt{2\omega N}(x-y)},
\ee
which is Eq. (12) in the main text.
It is also clear that this is nothing but the Fourier transform of $\langle \Psi_0| \hat n_k|\Psi_0\rangle$
obtained in Eq. (14) in the main text from the GGE, thus showing that infinite time limit and 
GGE results for $C(x,y)$ are equal.
 
\subsection{The two-point bosonic correlation function}
The impenetrable bosons field operators $\hat{\Phi}(x)$  are related to the fermionic ones $\hat{\Psi}(x)$ by the 
Jordan-Wigner transformation
\begin{eqnarray}\label{JordanWigner}
\hat{\Psi}(x) & = & \exp\left\{i\pi\int_{0}^{x}dz\,\hat{\Psi}^{\dag}(z) \hat{\Psi}(z)\right\}\, \hat{\Phi}(x),\qquad
\hat{\Psi}^{\dag}(x)  =  \hat{\Phi}^{\dag}(x)\,\exp\left\{-i\pi\int_{0}^{x}dz\,\hat{\Psi}^{\dag}(z) \hat{\Psi}(z)\right\}.\nonumber
\end{eqnarray}
Consequently, the equal time two-point bosonic correlation function  $C_{B}(x,y;t)$ with $y>x$ is 
\begin{equation}
C_{B}(x,y;t) \equiv \langle  \hat{\Phi}^{\dag}(x) \hat{\Phi}(y)\rangle
= \left\langle  \hat{\Psi}^{\dag}(x)\,\exp\left\{-i\pi\int_{x}^{y}dz\,\hat{\Psi}^{\dag}(z) \hat{\Psi}(z)\right\}  \hat{\Psi}(y) \right\rangle.
\end{equation}
Expanding the exponential and using Wick theorem one has
\bea
C_{B}(x,y;t) & = & \sum_{n=0}^{\infty}\frac{(-i\pi)^n}{n!}\int_{x}^{y}dz_{1}\cdots\int_{x}^{y}dz_{n}
\langle \hat{\Psi}^{\dag}(x) \hat{\Psi}^{\dag}(z_{1})\hat{\Psi}(z_1)\cdots  \hat{\Psi}^{\dag}(z_{n})\hat{\Psi}(z_n)  \hat{\Psi}(y)\rangle
\label{Cdet}
\\ \nonumber
& = &  \sum_{n=0}^{\infty}\frac{(-2)^n}{n!}\int_{x}^{y}dz_{1}\cdots\int_{x}^{y}dz_{n} \, \det_{ij} C(x_i,y_j;t),
\eea
where the indices $i,j$ run from $0$ to $n$, and we fixed $x_{i}=y_{i}\equiv z_{i},\,\forall i>0$, and $x_{0}\equiv x,\,y_{0}\equiv y$. 
The previous equation  is a Fredholm's minor of the first order. 

In  order to evaluate these correlators it is more convenient to introduce the $N\times N$ overlap matrix 
$\mathbb{A}(x,y;t)$ with elements 
\be
{\mathbb A}_{ij}(x,y;t) \equiv \int_{x}^{y}dz\,\Phi^{*}_{i}(z,t)\Phi_{j}(z,t),\quad i,j\in[0,\ldots,N-1].
\ee
in terms of which one has  \cite{pezer} 
\be\label{C_B_overlap}
C_{B}(x,y;t) = \sum_{i,j=0}^{N-1}\Phi^{*}_{i}(x,t) \mathbb{B}_{ij}(x,y;t) \Phi_{j}(y,t),
\ee
where the $N\times N$ matrix $\mathbb{B}(x,y;t)$ is
\be
\mathbb{B}(x,y;t) \equiv \det[\mathbb{P}](\mathbb{P}^{-1})^{T}, \qquad {\rm with } \quad
\mathbb{P}(x,y;t) \equiv \mathbb{I} -2\,\mathrm{sgn}(y-x)\mathbb{A}(x,y;t),
\ee
and $\mathbb{I}$ is the $N\times N$ identity matrix.

\paragraph{The large-time limit.}
Eq. (\ref{C_B_overlap}) is a good starting point to evaluate analytically the large-time limit of the bosonic correlation function. 
Indeed, the one-particle time-evolved functions are given in Eqs. (6) and (7) of the main text, and proceeding as in Eq. 
(\ref{Cxyt_1}) we can write  in the TD and large-time limits
$\Phi^{*}_{a}(z,t)\Phi_{b}(z,t)$ as
\bea
\Phi^{*}_{a}(z,t)\Phi_{b}(z,t) & \simeq & 
\frac{i^{a-b}}{\omega t} \sum_{p=-\infty}^{\infty} \chi_{a}^{*}\left(\frac{z + pL}{\omega t}\right)\chi_{b}\left(\frac{z + pL}{\omega t}\right) 
\simeq \frac{i^{a-b}}{L}\int_{-\infty}^{\infty} dx \chi_{a}^{*}\left(\frac{z}{\omega t}+x\right) \chi_{b}\left(\frac{z}{\omega t}+x\right) 
= \frac{1}{L}\delta_{ab},
\eea
where in the last equality we used the orthonormality of the eigenfunctions $\chi_{a}(x)$. 
Consequently, the large-time behavior of the $\mathbb{A}$ and $\mathbb{P}$ matrices is
\begin{equation}
\mathbb{A}_{ab}\equiv  \int_{x}^{y}dz\,\Phi^{*}_{a}(z,t)\Phi_{b}(z,t)= \frac{y-x}L \delta_{ab},\qquad
\mathbb{P}(x,y;t\rightarrow \infty) = \left(1-2\frac{|x-y|}{L}\right)\mathbb{I}.
\label{AandP}
\end{equation}
Clearly the above equations are valid as long as the rhs' are finite, i.e. when $|x-y|/L\sim O(1)$.
For $|x-y|\ll L$ different approaches must be used, as e.g. expanding the determinant in Eq. (\ref{Cdet}).

From Eq. (\ref{AandP}), the $\mathbb B$ matrix is 
\begin{equation}\label{Bmatrix_larget}
\mathbb{B} (x,y;t\rightarrow\infty) =  \left(1-2\frac{|x-y|}{L}\right)^{N-1} \mathbb{I},\quad {\rm and}
\quad \lim_{N\to\infty}  \mathbb{B} (x,y;t\rightarrow\infty)= \mathbb{I}\, \mathrm{e}^{-2n|x-y|} ,
\end{equation}
where the large $N$ limit has been taken keeping, as usual, $n=N/L$ constant.
Substituting Eq. (\ref{Bmatrix_larget}) in Eq. (\ref{C_B_overlap}), we finally have Eq. (16) in the main text, i.e. 
\begin{eqnarray}
C_{B}(x,y;t\rightarrow\infty) & = & C(x,y;t\rightarrow\infty)\mathrm{e}^{-2n|x-y|}. 
\label{CBinf}
\end{eqnarray}

\begin{figure}[t]
\includegraphics[width=0.5\textwidth]{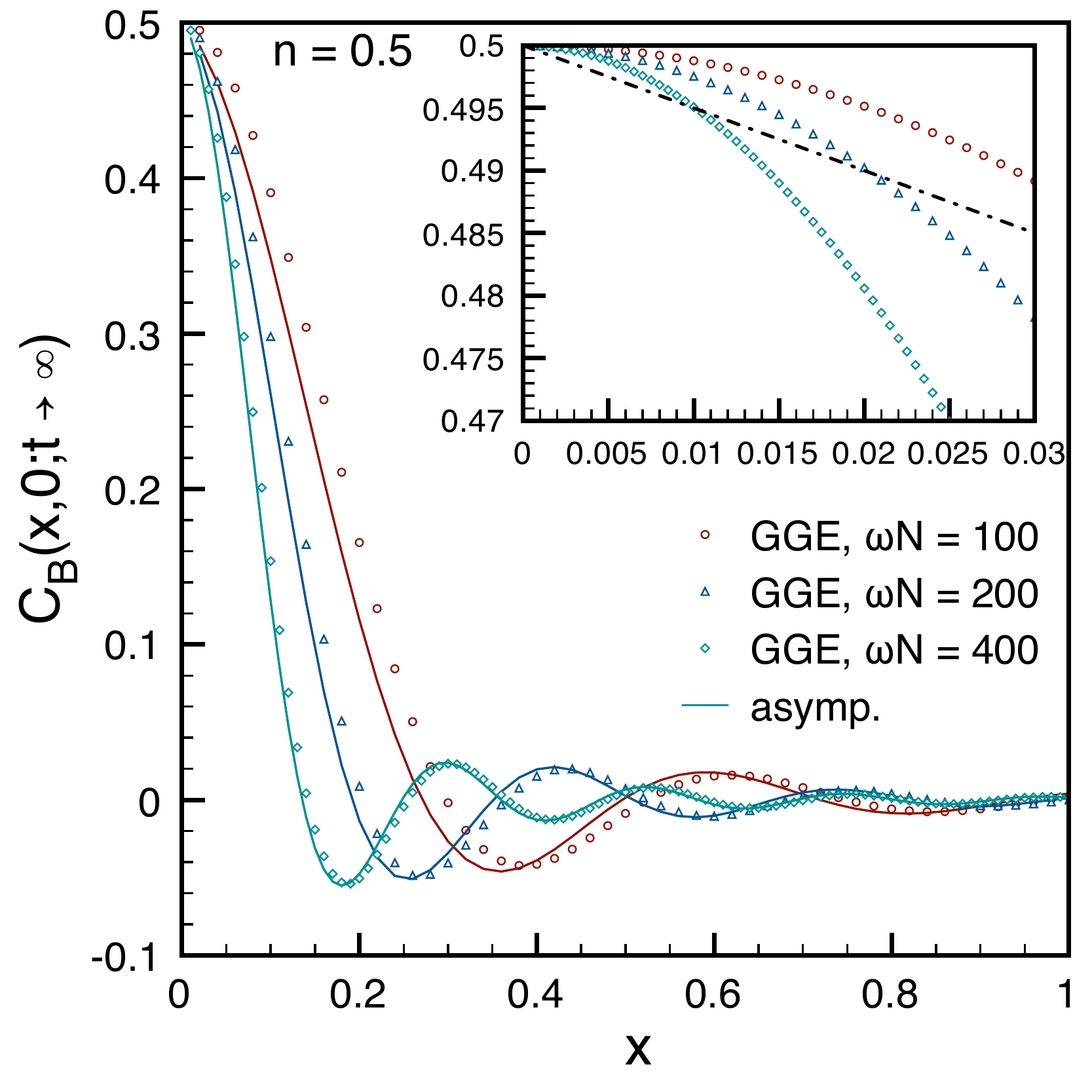}
\caption{Exact bosonic correlation function $C_{B}(x,y;t\rightarrow\infty)$ calculated by discretizing the Fredholm's 
minor in Eq. (\ref{Cdet}). For large enough $x$, the data always agree with the prediction in Eq. (\ref{CBinf}) (full lines), 
while for smaller $x$, the data approach it only for large enough $\omega N$.
The inset shows a zoom for very small $x$, for which the asymptotic $|x|$ behavior (dashed line) 
crosses over to a standard (non-singular) quadratic form.
} 
\label{newfig}
\end{figure}

As anticipated in the main text, Eq. (\ref{CBinf}) is valid only in the TD limit in the regime $|x-y|/L\sim O(1)$. 
For $|x-y|\ll L$,  the correlation function $C_{B}(x,y;t\rightarrow\infty)$ crosses over to the standard singular 
behavior $|x-y|^3$, as expected from general arguments. 
In order to show the correctness of this statement, we calculate numerically $C_{B}(x,0;t\to\infty)$
by discretizing the  Fredholm's minor in Eq. (\ref{Cdet}) as explained in Ref. \cite{adi} and using as input the 
GGE fermion correlation in Eq. (12) of the main text.
In Fig. \ref{newfig}, we report the numerically calculated $C_{B}(x,0;t\to\infty)$ as function of $x$ for different values of $\omega N$
(we recall $n=N/L$ is constant).
It is clear that increasing $N$, the numerical data approach the asymptotic result in Eq. (\ref{CBinf}).
However, if we zoom in the region of very small distances, as done in the inset of Fig. \ref{newfig}, the $|x-y|$ singularity is absent, 
as expected, and the main singularity is of the form  $|x-y|^3$ while the leading behavior is non-singular $(x-y)^2$.

\end{document}